\documentstyle[multicol,epsf,aps]{revtex}
\begin{document}
\def\a{\alpha}
\def\b{\beta}
\def\g{\gamma}
\def\F{\phi}
\def\T{\theta}
\def\D{\Delta}
\def\p{\pi}
\def\k{\xi}
\def\e{\epsilon}
\def\ka{\sigma}
\def\om{\omega}
\def\m{\mu}
\def\r{\rho}
\def\z{\zeta}
\def\d{\delta}

\title{Correlation functions in decorated lattice models
}

\author{I. Ispolatov$^*$, K. Koga$^{\dagger}$, and B. Widom$^*$}
\address{$^*$Department of Chemistry, Baker Laboratory, 
Cornell University, Ithaca,
NY 14853, USA.
\linebreak 
$^{\dagger}$Department of Chemistry, Fukuoka University of Education,
Munakata, Fukuoka 811-4192, Japan.
} 
\maketitle

\begin{abstract} 
\noindent  
Occupation probabilities for primary-secondary-primary cell strings and
correlation functions for primary sites of a decorated lattice model are expressed 
through the well-studied partition function and 
correlation functions of the Ising model. The results are analogous to those
found in related lattice models of
hydrophobic interactions and are interpreted in similar terms.

\end{abstract}
\section{Introduction}

Decorated-lattice-gas models that are equivalent to an underlying Ising 
model have been important as models of two-component mixtures with closed-loop
solubility curves \cite{m,wh1,wh2}. In summing over the states of the 
decoration sites in the partition function one obtains an Ising
model in which the energy and field parameters are related to the parameters
of the original mixture model by certain rules (transcription relations).
The interactions in the Ising model or in the equivalent one-component
lattice gas may then be understood to have arisen, or to have been altered, 
through the mediation of the particles occupying the decorated sites (cells).
In particular, the correlation functions between the particles occupying 
the primary cells, which would be those of the underlying Ising model or 
one-component lattice gas, may be understood to have resulted from such 
mediation. This is then analogous to the circumstance in recent lattice models
of hydrophobic interactions \cite{kw,bw,us}, in which the solvent-mediated
potential of mean force between solute molecules is obtained in terms of the 
correlation functions of the pure solvent. In the present work the potentials
of mean force between occupants of the primary cells in a class of 
decorated-lattice-gas models with closed-loop solubility curves 
\cite{m,wh1,wh2} are calculated and interpreted in terms similar to those in the lattice models of hydrophobic interaction \cite{kw,bw,us}.

The model and the calculational machinery are outlined in the next section and the results are then illustrated by numerical examples in Section III.
In Section IV the correlation lengths, which are the exponential decay lengths
of the various correlation functions, are calculated from those of the underlying Ising model and are displayed numerically.

\section{Decorated-lattice-gas model}
We start with a brief review of a  version of 
{\it Mermin's decorated lattice model} \cite{m}, Fig.~1, 
that describes a liquid mixture that possesses upper and lower critical 
solution points
and a closed-loop temperature-concentration coexistence curve \cite{wh1,wh2}.
\begin{figure}[b]
\centerline{\epsfxsize=6cm \epsfbox{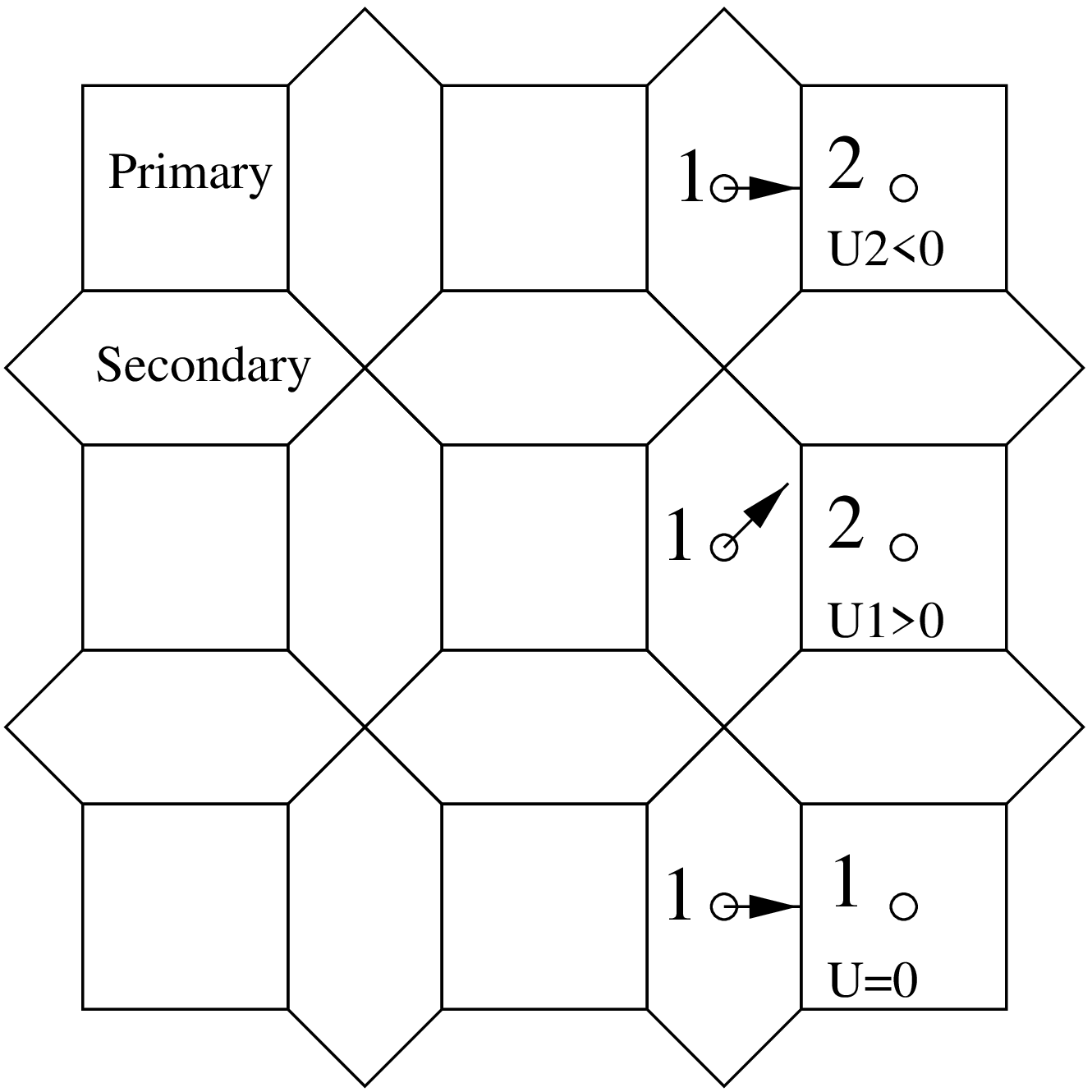}}
\noindent
{\small {\bf Fig.~1.}
Mermin's decorated lattice model
}
\end{figure}
Each cell, primary or 
secondary, is occupied by a molecule, either of type 1 or 2. 
Each molecule of either 
type has
$\om$ possible orientations. The only interaction takes place
between the molecules in 
adjacent primary-secondary cells. This interaction is defined to be 0 unless
adjacent primary-secondary cells are occupied by molecules of
different types, and its value then depends only on the orientation
of the molecule in the secondary cell. For each primary-secondary 
neighboring pair, 
if the occupant of the secondary cell points
to the primary cell occupied by an unlike molecule, the energy of interaction is
$U_2<0$; if it points in any other of $\om-1$ directions, the energy is $U_1>0$.

To calculate the partition function for the whole system 
we first write the partition functions $Q_{ijk}$ for all 
{\it primary-secondary-primary cell} strings for fixed occupations $i$ and $k$
of primary cells,
\begin {eqnarray}
\label{Q}
\nonumber
& Q_{111}=Q_{222}=\om&\\
& Q_{121}=Q_{212}=2 e^{-(U_1+U_2)\over kT} + (\om-2) e^{-2U_1\over kT}&\\
& Q_{122}=Q_{211}=Q_{112}=Q_{221}=e^{-U_2\over kT} + (\om-1) e^{-U_1\over kT}.&
\nonumber
\end{eqnarray} 
We also introduce the activity ratio $\z=z_1/z_2$
\begin{equation}
\label{z}
\z= e^{\m_1-\m_2 \over kT}[{ m_1\over m_2 }]^{d\over 2}, 
\end{equation}
where $m_i$ and $\m_i$ are molecular masses and chemical potentials 
for each species and $d$ is the dimensionality of space.

S in Eq.~(62) of Ref.~\cite{wh1}, 
we express the partition function of 
the model for the fixed number $N$ of primary cells as
\begin {eqnarray}
\label{Y}
& Y=\om^N \displaystyle \sum_{\{n_i\}} \z^{N_1} 
(Q_{111}\z+Q_{121})^{N_{11}}&\\
\nonumber
&(Q_{212}\z+Q_{222})^{N_{22}}
(Q_{112}\z+Q_{122})^{N_{12}}.& 
\end{eqnarray} 
Here 
$N_{i}$ is the number of molecules of component $i$,
$N_{ij}$ is the number of $i,j$ pairs of neighboring primary cells, and
the sum is over all possible occupations ``$\{n_i\}$'' of the primary cells.

Now we can evaluate the probability $P_{ijk}$ to find a
primary-secondary-primary cell string being occupied by the molecules
$i$, $j$, and $k$. It can be written as
\begin {eqnarray}
\label{ijk}
& P_{ijk}={2\over qN}{1\over Y}\om^N \displaystyle \sum_{\{n_i\}} \z^{N_1} 
{Q_{ijk}[1+\d_{1j}(\z-1)] \over \sum_{n=1}^2 Q_{ink}[1+\d_{1j}(\z-1)]}
N_{ik}&\\
\nonumber
&(Q_{111}\z+Q_{121})^{N_{11}}
(Q_{212}\z+Q_{222})^{N_{22}}
(Q_{112}\z+Q_{122})^{N_{12}}.&
\nonumber 
\end{eqnarray}
Here ${qN \over 2 }$ is the total number of bonds between $N$ primary cells of 
the lattice with coordination number $q$.
If we multiply $Q_{ijk}$ by an auxiliary factor $\a_{ijk}$, so that 
$Q_{ijk}$ is replaced by $Q_{ijk} \a_{ijk}$, the probability $P_{ijk}$ 
can be expressed as
\begin {equation}
\label{ijk2}
P_{ijk}={2\over qN}\left. 
{\partial \ln Y \over \partial \a_{ijk}}\right|_{\a_{ijk}=0}
\end{equation} 

We also can calculate the number of occupants of secondary cells pointing towards
the opposite species occupants of primary cells, $N_{p}$. This quantity is a 
measure of orientational ordering of the system. Note that the orientations of 
molecules in primary cells do not enter the expression for the partition 
function and therefore are isotropic everywhere in the phase diagram.
Since every secondary-primary cell neighboring pair where the cells are occupied by
opposite species and secondary cell occupants point to the primary cell comes 
with a factor $e^{-U_2 \over kT}$, there is no need to introduce any auxiliary 
multipliers:

\begin {equation}
\label{Nor}
N_{p}=-{1\over kT} 
{\partial \ln Y \over \partial U_2}
\end{equation} 

The structure of this expression is similar to that of 
(\ref{ijk}); it gives the 
probability of finding a particular ordered primary-secondary-primary string
multiplied by the average number of such strings.

To get the corresponding intensive quantity $n_{p}$, we need to divide
$N_{p}$ by the number of primary-secondary cell neighboring pairs 
$qN$ (twice the 
number of 
primary-primary neighboring pairs), $n_{p}=N_{p}/qN$.

Another important class of correlation functions it is possible 
to calculate is a connected pair and higher-order
correlation functions for the primary cells. 
To underline the connection to the Ising model, 
let us introduce the primary cell
occupation numbers, $n_j=+1$ for the $j$th cell being occupied by a molecule 
of species 1
and $n_j=-1$ if it is occupied by a molecule of species 2.
Then $N_1$ and $N_2$ in Eq.~(\ref{Y}) can be expressed as
$N_1=(N+\sum_{j=1}^N n_j)/2$ and $N_2=(N-\sum_{j=1}^N n_j)/2$.
We also introduce an auxiliary local field $\g_j$, in the presence of which the 
partition function takes the form:

\begin {eqnarray}
\label{g}
& Y=\om^N \z^{N/2}\displaystyle \sum_{\{n_i\}} e^{{1\over 2}
\sum_{i=1}^N n_i(\g_i+{1\over 2}\ln{\z})}&\\
&A_{11}^{N_{11}}
A_{22}^{N_{22}}
A_{12}^{N_{12}}.& 
\nonumber 
\end{eqnarray}
Here we used the shorthand notations: $A_{11}\equiv Q_{111}\z+Q_{121}$,
$A_{22} \equiv Q_{212}\z+Q_{222}$,  $A_{12}\equiv Q_{112}\z+Q_{122}$.

As in the Ising model, the two-body connected correlation function for the
primary cells $i$ and $j$, $C_{i,j}\equiv \langle n_i n_j\rangle  - \langle n_i\rangle \langle n_j\rangle $, can be 
expressed as
\begin {equation}
\label{corr}
C_{i,j}=\left. {\partial^2 \ln Y \over \partial \g_i \partial \g_j}
\right|_{\{\g_k\}=0}. \:\:
\end{equation} 
In order to complete the calculation of correlation functions $P_{ijk}$ and
$C_{ij}$, we express the partition function $Y(\{\g_i\})$ through the 
known partition and correlation functions for the Ising model.
Using lattice identities, $q N_1=2N_{11}+N_{12}$ and $q N_2= 2N_{22}+N_{12}$, 
we reduce Eq.~(\ref{g}) to
\begin {eqnarray}
\label{gm}
& Y=\om^N \z^{N/2} (A_{11} A_{22})^{qN\over 4 }
\displaystyle \sum_{\{n_i\}} e^{{1\over 2}
\sum_{i=1}^N n_i(\g_i+{1\over 2}\ln{\z}+{q\over 4} 
\ln{A_{11}\over A_{22}})}&\\
&[{A_{12}\over \sqrt{A_{11}A_{22}}}]^{N_{12}}.& 
\nonumber 
\end{eqnarray}
On the other hand, for the Ising model with the Hamiltonian
\begin{equation}
\label{ising1}
{\cal H}=- J \displaystyle \sum_{\langle i,j\rangle }s_i s_j - \displaystyle \sum_{i} s_i(H+\g_i),
\end{equation}
the partition function $Z$ and the connected two-point correlation function 
$C_{i,j}^{Ising}$ can be expressed as
\begin{equation}
\label{ising2}
Z=e^{-{qN\over 2}} \displaystyle \sum_{\{s_i\}}e^{
{1\over kT}\sum_{i=1}^N n_i(H+\g_i)+2JN_{+-} },
\end{equation}
\begin {equation}
\label{corri}
C_{i,j}^{Ising}=\left.{\partial^2 \ln Z \over \partial \g_i \partial \g_j}
\right|_{\{\g_k\}=0}. 
\end{equation} 

Here $N_{+-}$ is the number of bonds between up and down spins,
which is equivalent in the mixture model to the number of 1-2 neighboring pairs 
of primary cells,
$N_{+-}=N_{12}$.

We observe that the connected correlation function for the decorated
lattice model and Ising model are identical if the coupling constant $J$
and external field $H$ in the Ising model are defined as 
\begin {eqnarray}
\label{main}
&C_{i,j}=C_{i,j}^{Ising}&\\
&{J\over kT} <=> - {1\over 2} \ln [{A_{12}\over \sqrt{A_{11} A_{22}}}],\;  \; 
{H\over kT} <=> {1\over 2} \ln \z + {q\over 4} \ln{A_{11}\over A_{22}}.&
\nonumber
\end{eqnarray} 

We next proceed to the calculation of the $k$-point 
potential of mean force between the 
occupants of the
primary cells, $W_k(\{i \ldots j\})=-kT 
\ln[{\langle  \r_i \ldots  \r_j \rangle  \over \langle \r\rangle ^k}]$.
Since the effective interaction between occupants of primary cells 
depends on the species and state of the occupants of the intervening 
secondary cells, the corresponding primary-primary cell correlation functions 
may be viewed as having been mediated by those secondary-cell occupants.
This is analogous to the solvent-mediated correlations between solute
molecules in related hydrophobic-interaction models \cite{kw,bw,us},
but it is to be understood that in the present model the effective 
correlations are not those between molecules of the same species since 
molecules of both species may be the occupants of either kind of site.
Indeed, the correlations calculated here are those between the occupants
(whatever their species) of the same kind of site (the primary cells).

In the definition of $W$, $\r_i$ is the non-negative occupation number 
(density) of the
$i$th cell. Since the actual amplitude of $\r$ in the definition of
$W$ cancels, it is convenient to let it vary between 0 and 1, i.e.
to define it as $\r_i\equiv {n_i+1 \over 2}$. It allows 
a $k$-point potential of mean force to be expressed through the $k$- 
and lower-order
connected correlation functions in the Ising model.
For example, for $k=2$,
\begin {equation}
\label{pot}
W_2(\{i,j\})=-kT \ln[{C_{i,j}\over (\langle n\rangle +1)^2} +1].
\end{equation}
Here $\langle n\rangle ={\partial \ln Z \over \partial H}$ is the average occupation 
number or magnetization
for the corresponding Ising model.
The correspondence between the parameters of the decorated lattice model and
Ising model is given by (\ref{main}). 
Similarly, one can calculate the higher-than-second order correlation
functions and potentials of mean force.

Finally we show how to calculate the primary-secondary cell 
and secondary-secondary cell correlation functions and 
potentials of mean force.
Taking into account Eqs.~(\ref{Q}), we calculate a probability $S_{\{i1j\}}$ to 
find a particle of species
1 in the secondary cell situated between primary cells occupied by molecules 
of species $i$ and $j$:
\begin {eqnarray}
\label{pr}
\nonumber
&S_{\{111\}}={\om\over \om+2 e^{-(U_1+U_2)\over kT} + 
(\om-2) e^{-2U_1\over kT}}, &\\ 
&S_{\{112\}}=S_{\{122\}}={1\over 2},&\\ 
&S_{\{212\}}={2 e^{-(U_1+U_2)\over kT} + 
(\om-2) e^{-2U_1\over kT}\over \om+2 e^{-(U_1+U_2)\over kT} + 
(\om-2) e^{-2U_1\over kT}}.&\nonumber
\end{eqnarray}
Taking into account the definition of $\rho$ for primary cells given above, 
we express an average density of molecules of the species 1 in 
secondary cells, $\rho^{\dagger}$, as:
\begin {equation}
\label{rho}
\rho^{\dagger}=\langle \r_i\r_{j}\rangle  S_{\{111\}}+
2\langle (1-\r_i)\r_{j}\rangle  S_{\{211\}}+
\langle (1-\r_i)(1-\r_{j})\rangle  S_{\{212\}}.
\end {equation}
Here $\r_i$ and $\r_j$ are the occupation numbers of the
neighboring primary cells $i$ and $j$.
Similarly, the primary-secondary-cell correlation function 
$\langle  \r_k \r^{\dagger}_l\rangle $ 
is expressed through the following average:
\begin {eqnarray}
\label{prc}
&\langle  \r_k \r^{\dagger}_l\rangle =
\langle \r_k\r_i\r_{j}\rangle  S_{\{111\}}+
\langle \r_k(1-\r_i)\r_{j}\rangle  S_{\{211\}}+&\\
&\langle \r_k\r_i(1-\r_{j})\rangle  S_{\{112\}}+
\langle \r_k(1-\r_i)(1-\r_{j})\rangle  S_{\{212\}}.&
\nonumber
\end {eqnarray}
Here $i$ and $j$ are the primary cells between which the secondary
cell $l$ is situated.
It follows that the primary-secondary cell correlation function
of the decorated lattice model can be expressed through
the energy-magnetization (3-point) correlation function
of the underlying Ising model 
and various 2-point correlation functions.
Finally, for the primary-secondary cell potential of mean force
${\tilde W}_2(\{k,l\})$, one obtains:
\begin {equation}
\label{pmfrc} 
{\tilde W}_2(\{k,l\})=-kT \ln{\langle  \r_k \r^{\dagger}_l\rangle 
\over \langle  \r \rangle  \langle  \r^{\dagger}\rangle }.
\end{equation}. 
A similar approach shows that the secondary-secondary cell
correlation function and potential of mean force for the decorated lattice
model can be expressed through energy-energy (4-point), energy-magnetization
(3-point) and 2-point correlation functions.

Because of the generic nature of Mermin's decorated lattice model, these
results are applicable for lattices in arbitrary dimension $d$ with various
coordination numbers $q$. 

\section{Examples}
Using the formal mapping of the correlation functions of the decorated 
lattice model
onto the correlation functions of an Ising spin system, we perform computations
of the 2- and 3- body potential of mean force in two dimensions.
The following values were chosen for the 
model parameters: $U_1=-U_2$ (with $U_1$, unspecified, then a scale factor
for the temperature), $\om=100$, $\z=1$.
The phase diagram for these values of the parameters is shown in Fig.~2.
\begin{figure}
\centerline{\epsfxsize=8cm \epsfbox{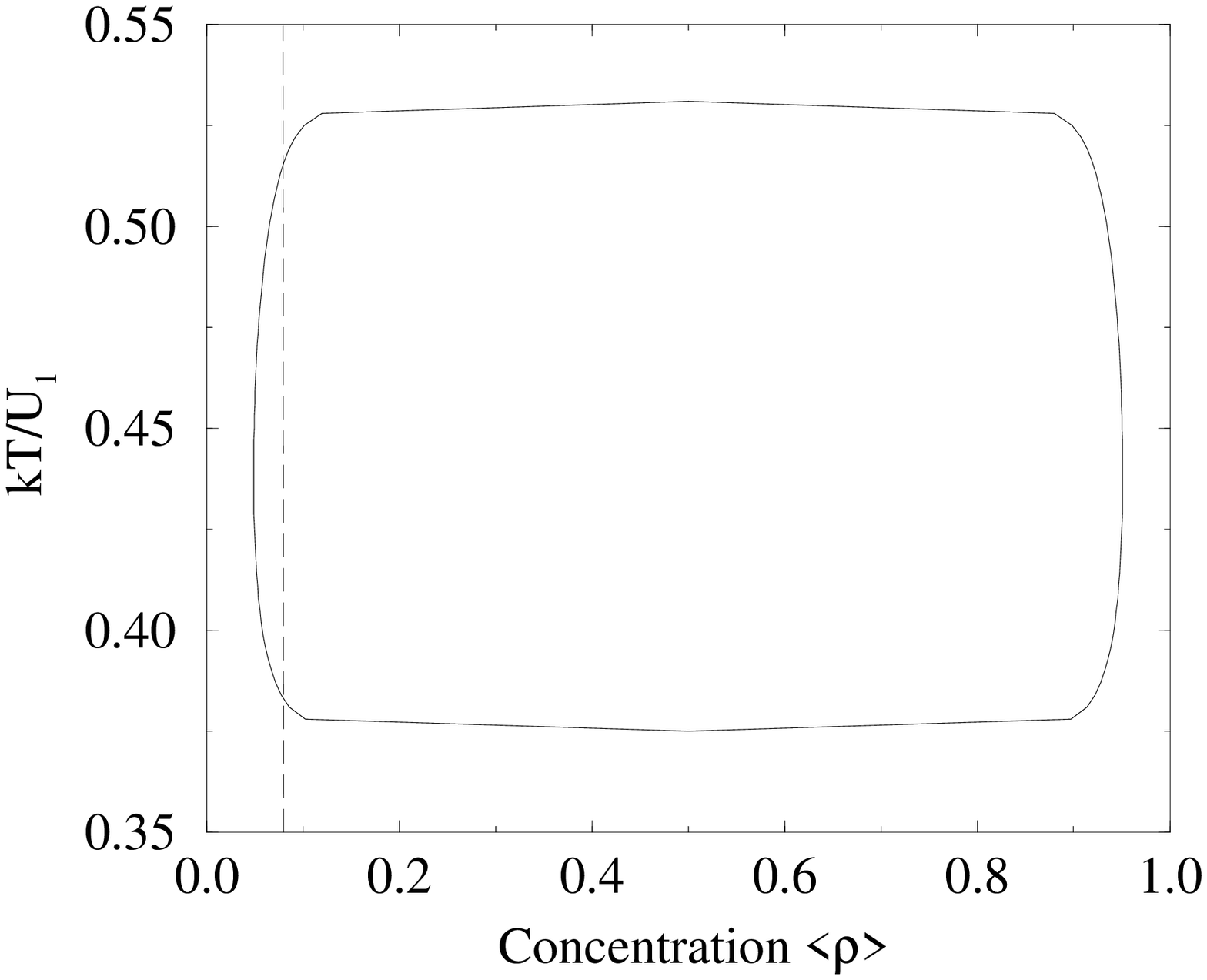}}
{\small {\bf Fig.~2.} 
Closed loop temperature-concentration solubility curve.}
The dashed line marks the concentration $\langle \rho \rangle  \approx 0.08$
\end{figure}

Despite the fact that the correlation functions for the Ising model are 
available analytically in the form of series expansions (see, for example, 
\cite{fisher}), 
we found it more convenient to
generate them each time directly using Monte Carlo simulations.
We used Metropolis and Wolff algorithms \cite{rg} 
with the lattice size varying from $100 \times 100 $
to $ 200 \times 200 $. All the results presented below were averaged over 
the whole system and over 200 configurations; configuration were considered
different if there were separated by about 10 flips for each spin
in the Wolff algorithm. The correlation functions and potentials of mean force 
were measured along the lattice (``normal'') unit vectors 
($(1,0)$ and $(0,1)$) and along 
the main diagonal $(1,1)$. We found that in general the results for
``normal'' and ``diagonal'' measurements are very similar.
An example is in Figure 3, where the two-body potential of mean force is
shown.
The dashed curve is a result of the superposition
of normal and diagonal plots, with normal and diagonal points occurring at 
integer and $\sqrt{2}\times$integer lattice separation.

The potentials of mean force $W_2$ and $W_3$ were calculated for the 
temperatures 
corresponding to 
the widest point of the phase diagram ($\langle \rho\rangle  \approx 0.05$, 
$kT/U_1 \approx 0.44$), 
and also for the two temperatures corresponding to the intersections 
of the dashed line
and the coexistence curve in Fig.~2, at both of which there is the same  
minority species concentration $\langle \rho\rangle   \approx 0.08$ 
($kT/U_1 \approx 0.384$ and $kT/U_1 \approx 0.513$).
The effective Ising coupling constant is $J/kT=0.47$ for the first point
and $J/kT=0.45$ for the next two.
\begin{figure}
\centerline{\epsfxsize=8cm \epsfbox{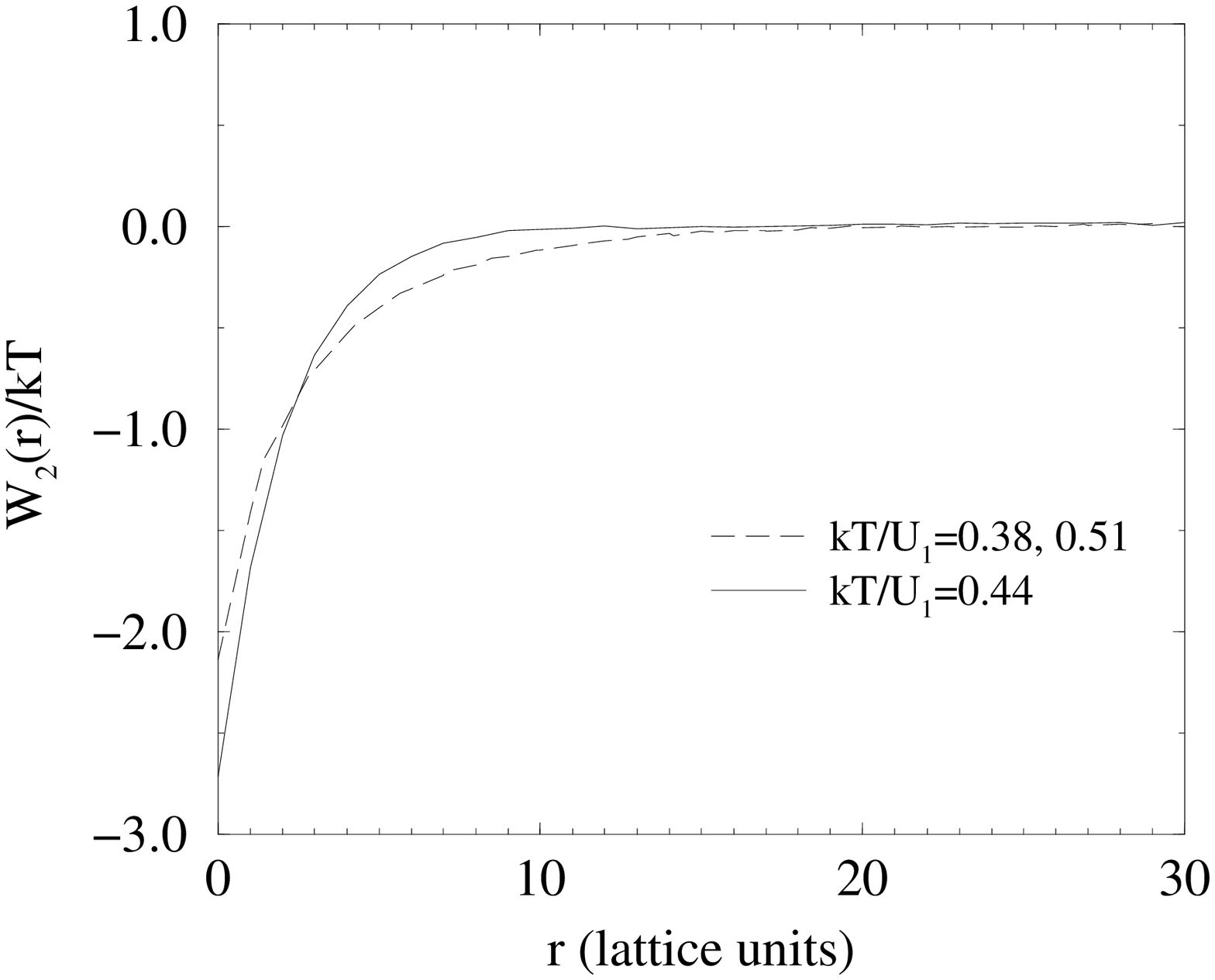}}
{\small {\bf Fig.~3.} 
Plots of the two-body potential of mean force at
$kT/U_1=0.44$ (solid line) and $kT/U_1=0.38, 0.51$ (dashed line).}
\end{figure}

The potentials of mean force for primary-secondary cell occupants
is shown in Fig.~4. The original primary cell and two primary cells
that surround the secondary cell lie on the same lattice vector;
the distance between primary and secondary cell is assumed to be equal to
half-sum of the distances between the original and two neighboring primary cells.

\begin{figure}
\centerline{\epsfxsize=8cm \epsfbox{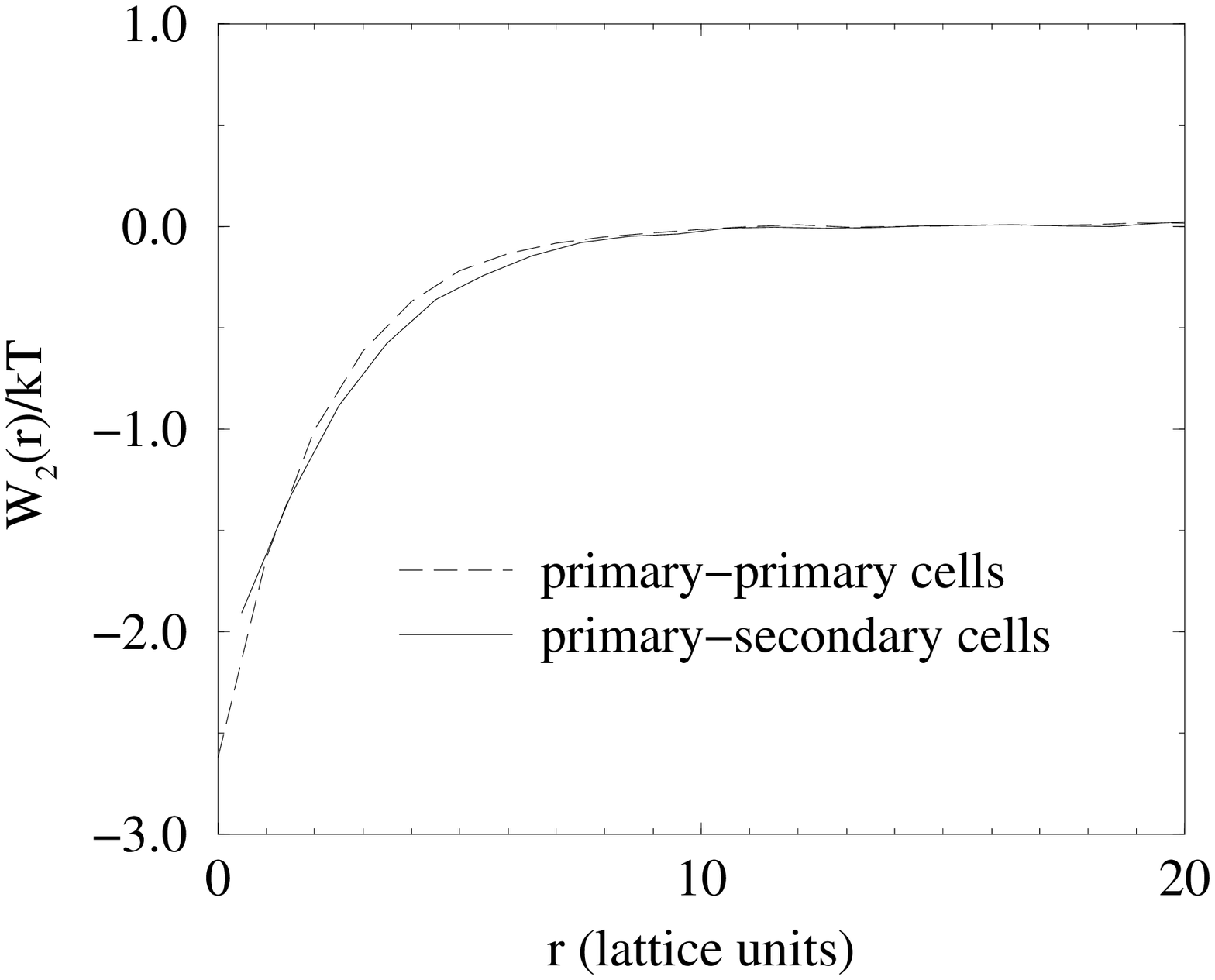}}
{\small {\bf Fig.~4.} 
Plots of the two-body potential of mean force at
$kT/U_1=0.44$ for primary-secondary cells (solid line)
and primary-primary cells (dashed line).} 
\end{figure}

For the three-body potential, we chose the two simplest configurations:
In the first one,
particles 1, 2, and 3 lie on a straight line with the distance $r$ between
particles 1 and 2 being equal to the distance between particles 2 and 3.
In the second configuration particles 1, 2, and 3 form a right triangle
in which the legs 1,2 and 1,3 are of equal length $r$.
We are mainly interested in the question of how well the effective
3-body interaction is approximated by the sum of the three pair
interactions.

In Fig.~5 we present a plot of the difference
between the sum of the three pair potentials
and the true 3-body potential for a linear configuration,
$\Delta W_3(r)\equiv2 W_2(r) + W_2(2r) - W_3(r,r,2r)$.
Since all potentials $W_2$ and $W_3$ are negative for any $r$,
one can observe that the sum of the three pair interactions
clearly overestimates (the absolute magnitude of) the true 3-body 
potential. 
However, unlike in
some one-dimensional models of hydrophobic interactions \cite{us},
this overestimate is less (in absolute value) than the interaction 
between the furthermost particles, 1 and 3. To show that, a plot
of the corresponding two-body potential $W_2 (2r)$ is presented in the same
Fig.~5.
\begin{figure}
\centerline{\epsfxsize=8cm \epsfbox{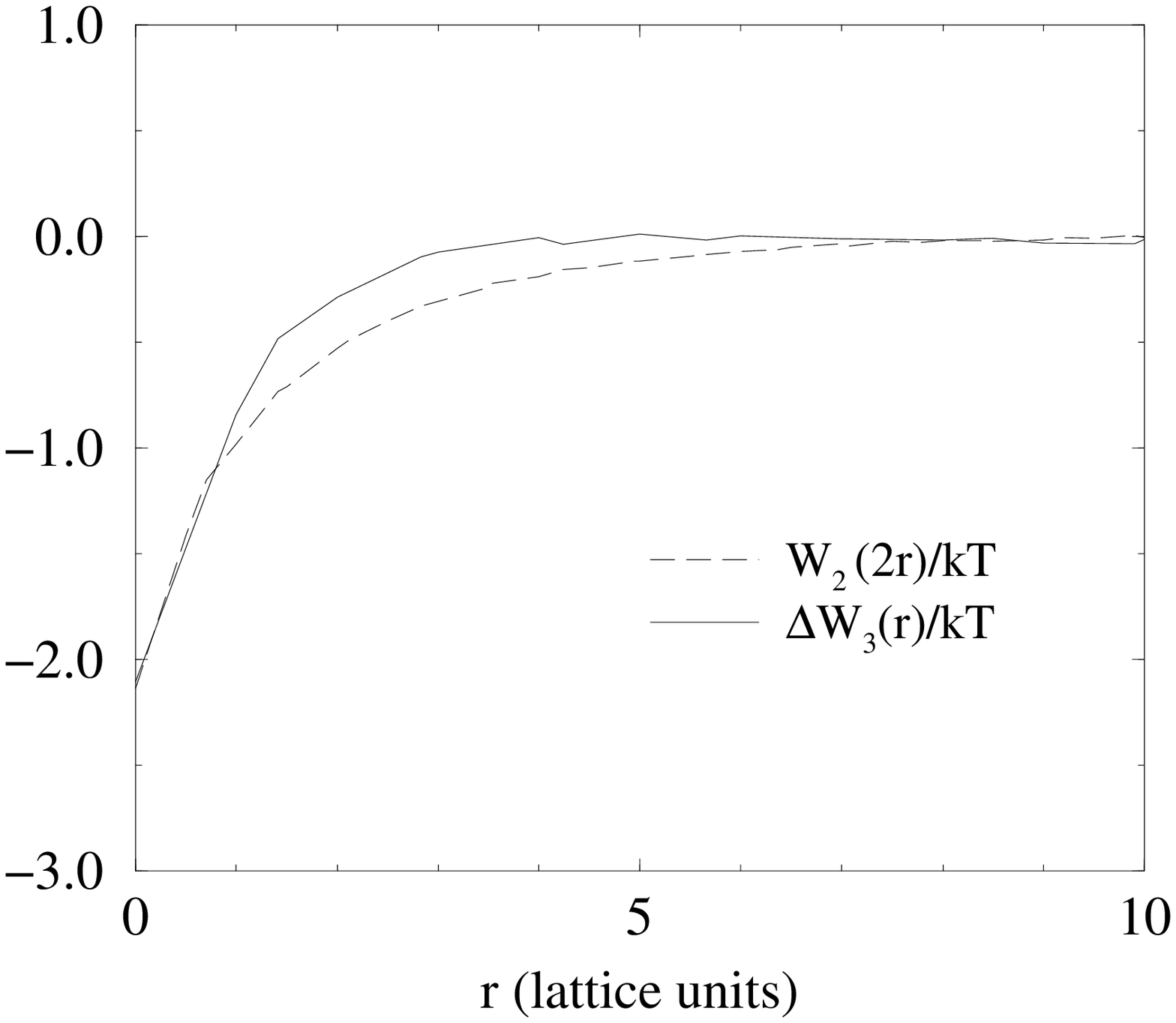}}
\noindent
{\small {\bf Fig.~5}.
Difference between the sum of the three pair
potentials and the true 
3-body interaction, $\Delta W_3(r)\equiv 2W_2(r) + W_2(2r) - W_3(r,r,2r)$
(solid line), and the pair potential $W_2(2r)$ (dashed line), 
for the linear 
configuration of points and the 
temperatures $kT/U_1=0.384$ or $0.513$
}
\end{figure}

Similar results for the triangular configuration of points are presented
in Fig.~6. Here again, as in the linear case, the sum of all pairwise 
interactions overestimates the true 3-body potential, but, as in the 
linear case, by less than the magnitude of the pairwise potential
between two most remote points, $W_2( \sqrt{2} r)$.

\begin{figure}
\centerline{\epsfxsize=8cm \epsfbox{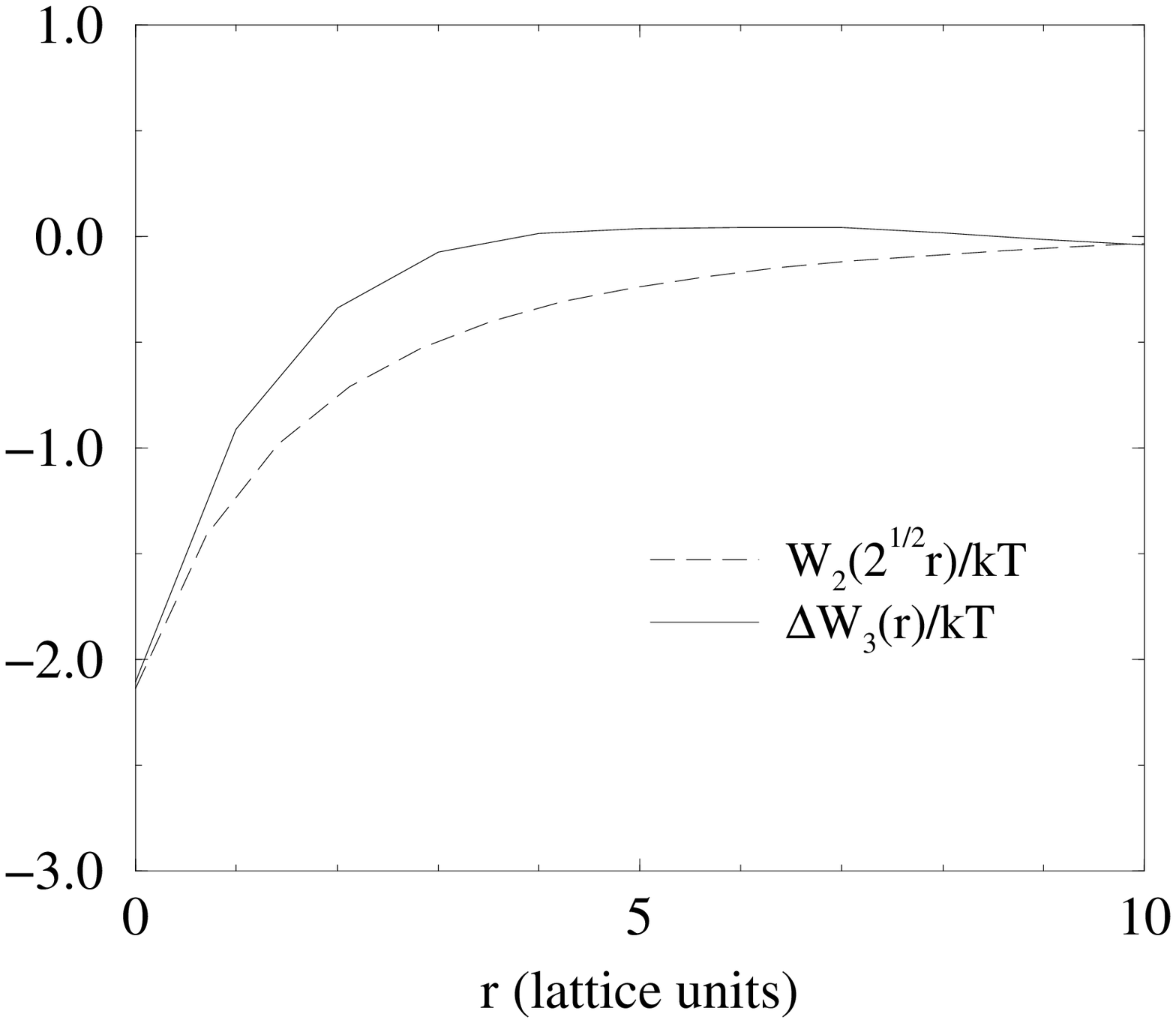}}
\noindent
{\small {\bf Fig.~6}.
Difference between the sum of the three pair
potentials and the true 
3-body interaction, $\Delta W_3(r) \equiv 2W_2(r) + W_2(\sqrt{2}r) 
- W_3(r,r,\sqrt{r})$
(solid line), and the pair potential $W_2(\sqrt{2}r)$
(dashed line), for the triangular 
configuration of points and the 
temperatures $kT/U_1=0.384$ or $0.513$
}
\end{figure}

\section{Correlation length}

In this section the correlation lengths for the decorated lattice model 
are calculated from those of the underlying Ising model. 
Before we proceed to the calculation, we note that 
the correlation lengths for any pair of species (i.e., 1-1, 1-2, 2-2) 
are all identical in this model.
The exponential range of correlation for a pair of 
molecules of species $\alpha$ and $\beta$ occupying the primary cells 
is defined by 
\begin{equation}
	1/\xi_{\alpha\beta} = - \lim_{|\vec{R}|\rightarrow \infty} 
            |\vec{R}|^{-1} \ln |h_{\alpha\beta}(\vec{R})|
\end{equation}
where $h_{\alpha\beta}(\vec{R})$ is the pair correlation function
for species $\alpha$ and $\beta$. However, all the pair correlation
functions $h_{11}$,$h_{12}$, and $h_{22}$ are expressed as
\begin{equation}
h_{\alpha\beta}(\vec{R}) = k_{\alpha\beta} C(\vec{R})
\end{equation}
where $k_{\alpha\beta}=(2\delta_{\alpha\beta}-1)/4\rho_\alpha\rho_\beta$
with $\rho_\alpha = N_\alpha/N$. Note that 
$C(\vec{R})$ is the two-body connected correlation function $C_{i,j}$
defined in 
Eq.~(\ref{corr}), 
but now expressed as a function of $\vec{R}=\vec{R_i}-\vec{R_j}$
instead of as a function of the primary cell addresses $i$,$j$.
Since the factor $k_{\alpha\beta}$ is independent of $\vec{R}$, 
all the correlation lengths for any pair of species 
in this model are identical to the length defined by
\begin{equation}
1/\xi = - \lim_{|\vec{R}|\rightarrow \infty} |\vec{R}|^{-1} 
\ln |C(\vec{R})|.
\label{eqn:20}
\end{equation}
On the other hand, the correlation length for the Ising model is defined by
\begin{equation}
1/\xi^{Ising} = - \lim_{|\vec{R}|\rightarrow \infty} |\vec{R}|^{-1} 
\ln |C^{Ising}(\vec{R})|.  
\end{equation}
Therefore the correlation length $\xi$ for the decorated lattice model is 
calculated from $\xi^{Ising}$ through the transcription given by 
Eq.~(\ref{main}). 

For the Ising model an accurate expression for 
the correlation length at zero field below $T_c$ is available
as a function of $\exp(-2J/kT)$\cite{fisher}.
Thus we can calculate $\xi$ for the decorated lattice model 
along the coexistence curve from the upper to the lower critical 
solution point. For the simple cubic lattice, 
the following expression is used: 
\begin{equation}
\frac{1}{\xi} = \frac{1}{f}{\cosh}^{-1}\left(
         1 + \frac{f^2}{2\Lambda^{\prime}_2}\right)
\label{eqn:23}
\end{equation}
or 
\begin{equation}
\frac{1}{\xi} = \frac{1}{f}\ln \left[ 1 + \frac{f^2}{2\Lambda^{\prime}_2}
+\sqrt{\left(1 + \frac{f^2}{2\Lambda^{\prime}_2}\right)^2-1}\right]
\label{eqn:24}
\end{equation}
where for the simple cubic lattice in the direction $\vec{\rm e}_0=(1,0,0)$, 
\begin{equation}
\Lambda^{\prime}_2(x)= x^4 -x^6 + 10x^8 -14 x^{10} 
+ 93 x^{12} -201x^{14}+ \cdots \quad {\rm and} \quad f^2=1
\label{eqn:25}
\end{equation}
and for the same lattice in the direction $\vec{\rm e}_1=(1,1,0)/\sqrt{2}$, 
\begin{equation}
\Lambda^{\prime}_2(x)= x^4 -\frac{3}{4}x^6 + 9\frac{7}{16}x^8 
-13\frac{11}{32}x^{10} +\cdots  \quad   {\rm and} \quad f^2=\frac{1}{2}.
\label{eqn:26}
\end{equation}
The variable $x$ in terms of which $\Lambda^{\prime}_2$ is expanded 
is to be understood as $A_{12}/\sqrt{A_{11}A_{22}}$ 
for the decorated lattice model, instead of $\exp(-2J/kT)$ for the Ising model.

Here we show the numerical results for the correlation length 
for the decorated lattice model of the simple cubic lattice. 
The results are obtained for various values for the parameter $\omega$
at $|U_2|/U_1=1$.
Shown in Fig.~7(a) is the behavior of the correlation length $\xi$
in the direction $\vec{\rm e}_0$ along 
the coexistence curves as a function of temperature 
between the upper and lower critical solution points. 
For any given $\omega$, 
the correlation length takes a minimum value at a temperature 
lower than the midpoint between the upper and lower critical temperatures.
With increasing $\omega$, the minimum of the curve 
becomes deeper and shifts toward the lower critical temperature. 
This means that a larger number of possible orientations $\omega$ 
gives rise to a correlation length that increases more rapidly as 
the system approaches the lower critical solution temperature.
Similar results are obtained for the direction $\vec{\rm e}_1$
as shown in Fig.~7(b).
\noindent
\begin{figure}
\centerline{\epsfxsize=7cm \epsfbox{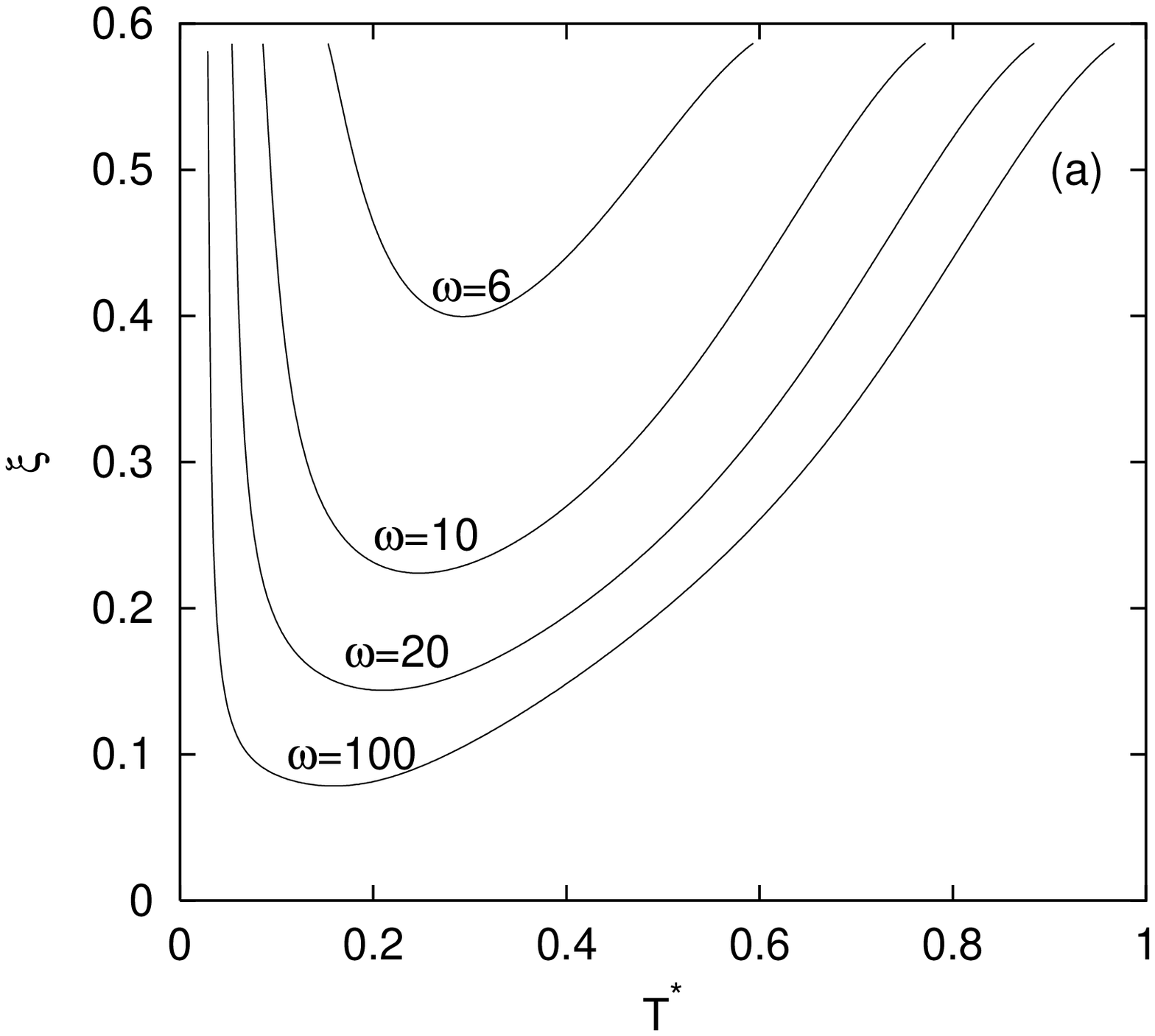}}
\centerline{\epsfxsize=7cm \epsfbox{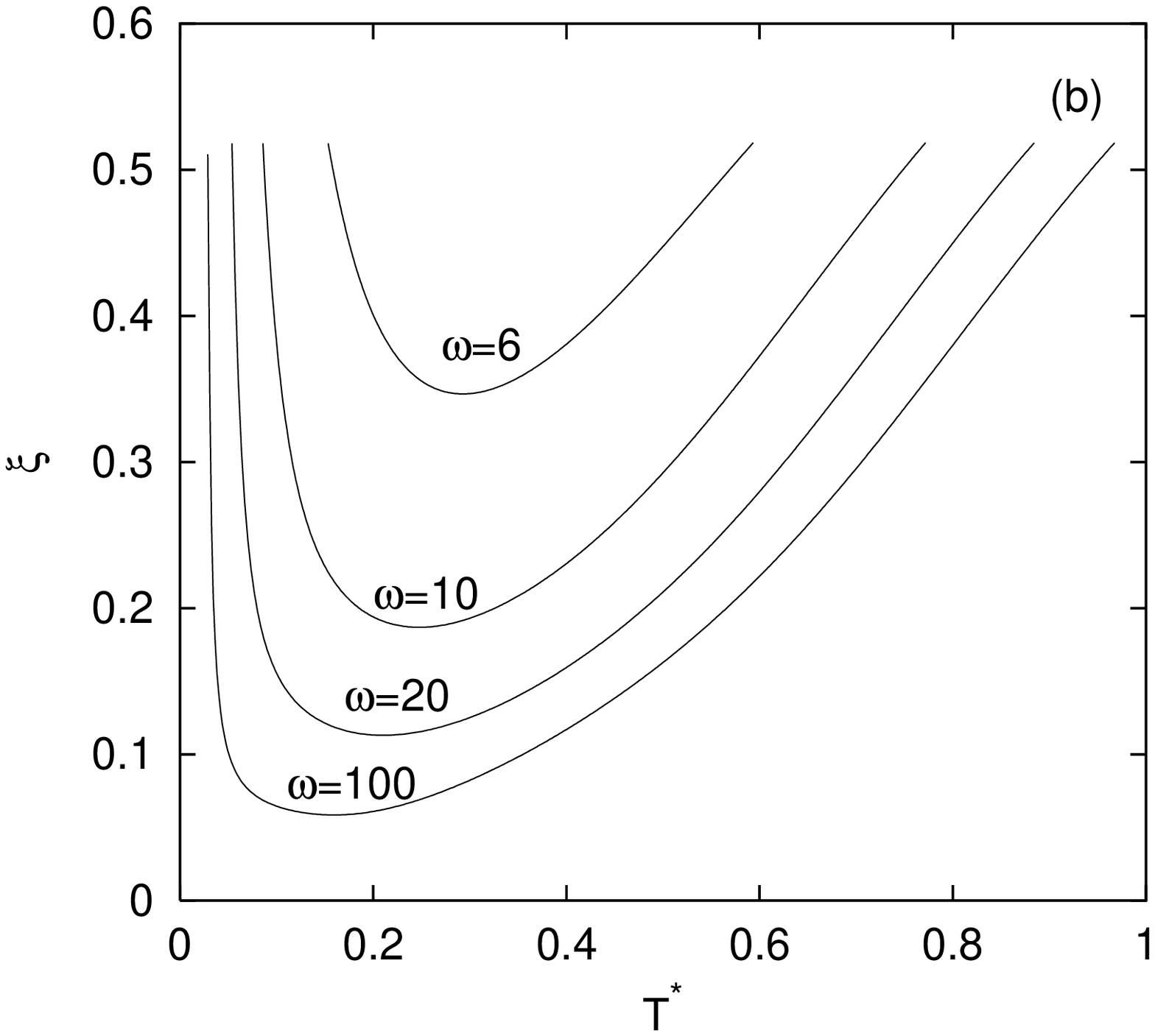}}
\noindent
{\small {\bf Fig.~7}.
 Behavior of the correlation length along the coexistence curve 
for the decorated lattice-gas model for various values of $\omega$ at 
$|U_2|/U_1=1$. The lattice type is the simple cubic and 
the lattice direction in which $\xi$ is calculated is (a) $\vec{e}_0=(1,0,0)$ 
and (b) $\vec{e}_1=(1,1,0)/\sqrt{2}$. Temperature is in units of $U_1/k$.
}
\end{figure}

\section{Summary}
We have calculated and illustrated
the correlations between primary cells
for a decorated-lattice-gas model
that exhibits 
a closed-loop phase diagram.
The three-body potential of mean force was also calculated and compared 
with 
the sum of the three two-body interactions. The calculations are related 
to those in 
earlier models of hydrophobic interactions and are interpreted in similar 
terms.

\section{Acknowledgments}
This work was supported by the U.S. National Science Foundation,
the  Cornell Center for Materials Research,
and by the Japan Society for the Promotion of Science.

\vfill
\eject

\end{document}